\documentclass[a4paper]{PoS}

\usepackage[utf8]{inputenc}
\usepackage[T1]{fontenc}

\usepackage{subcaption}
\usepackage[font=small,labelfont=bf]{caption}

\usepackage[separate-uncertainty = true]{siunitx}
\usepackage{amsmath}
\usepackage{booktabs}

\usepackage{microtype}

\DeclareSIUnit\gauss{G}
\newcommand{\EeV}{\exa\electronvolt}

\graphicspath{{img/}}

\let\OLDthebibliography\thebibliography
\renewcommand\thebibliography[1]{
  \OLDthebibliography{#1}
  \setlength{\parskip}{5pt}
  \setlength{\itemsep}{0pt plus 0.3ex}
}

\title{Modelling uncertainty of the radiation energy emitted by extensive air showers}

\ShortTitle{Modelling uncertainty of the radiation energy emitted by extensive air showers}

\author{Marvin Gottowik$^a$, Christian Glaser$^b$, \speaker{Tim Huege}$^c$, Julian Rautenberg$^a$%
        \\
        \llap{$^a$}Bergische Universität Wuppertal, Gaußstraße 20, 42097 Wuppertal, Germany\\
        \llap{$^b$}RWTH Aachen University, III. Physikalisches Institut A, Aachen, Germany\\
        \llap{$^c$}IKP, Karlsruher Institut fur Technologie, Postfach 3640, 76021 Karlsruhe, Germany\\
        E-mail: \email{gottowik@uni-wuppertal.de}, \email{glaser@physik.rwth-aachen.de}, \email{tim.huege@kit.edu},     \email{jrautenb@uni-wuppertal.de}}

\abstract{Recently, the energy determination of extensive air showers using radio emission has been shown to be both precise and accurate. In particular, radio detection offers the opportunity for an independent measurement of the absolute energy of cosmic rays, since the radiation energy (the energy radiated in the form of radio signals) can be predicted using first-principle calculations involving no free parameters, and the measurement of radio waves is not subject to any significant absorption or scattering in the atmosphere. Here, we verify the implementation of radiation-energy calculations from microscopic simulation codes by comparing Monte Carlo simulations made with the two codes CoREAS and ZHAireS. 
To isolate potential differences in the radio-emission calculation from differences in the air-shower simulation, the simulations are performed with equivalent settings, especially the same model for the hadronic interactions and the description of the atmosphere. Comparing a large set of simulations with different primary energies and shower 
directions we observe differences amounting to a total of only \SI{3.3}{\percent}. This corresponds to an uncertainty of only \SI{1.6}{\percent} in the determination of the absolute energy scale and thus opens the potential of using the radiation energy as an accurate calibration method for cosmic ray experiments.}

\FullConference{35th International Cosmic Ray Conference - ICRC217-\\
		10 -- 20 July, 2017\\
		Bexco, Busan, Korea}

\begin{document}

\section{Introduction}
When ultra high energy cosmic rays (UHECR) interact with the Earth‘s atmosphere an extensive air shower (EAS) is produced.
Several methods are feasible to detect such an EAS, like measurement of the footprint of the particles reaching the ground.
Additionally, the longitudinal profile can be analysed using the emitted fluorescence light.
A recent method is the detection via radio emission which is used e.g. at the Auger Engineering Radio Array (AERA) \cite{AERA} of the Pierre Auger Observatory \cite{PAO}.

Two mechanisms contribute to coherent radio emission from air showers. The dominant geomagnetic emission
induced  by  charged  particle  motion  in  the  Earth’s  magnetic  field $\vec{B}$ is polarized in the direction of the Lorentz force ($\vec{v} \times \vec{B}$) with shower direction denoted by $\vec{v}$.
The time-varying negative charge excess in the shower front is due to the knock-out of
electrons from air molecules and annihilation of positrons in the shower front and gives rise to radiation polarized radially towards the shower core. 
Instead of explicitly modelling the two emission mechanisms, the radio emission emitted by an extensive air shower 
can be calculated directly from the movement of the shower particles from first principles using classical electrodynamics \cite{Huege:2016veh}.

The emission originates from the well-understood electromagnetic part of the air shower. 
Hence, the radiation energy can be used as an estimator for the cosmic-ray energy which is already successfully exploited by AERA \cite{Aab:2015vta, Aab:2016eeq}. 
Here, we estimate the accuracy of the calculation of the radiation energy by comparing the prediction of the two independent air-shower simulation codes CoREAS \cite{CoREAS} and ZHAireS \cite{ZHAireS}.

\section{Simulation Setup}
For the simulation of the air showers, two different programs CORSIKA 7.4100 and Aires 2.8.4a are used.
For the computation of the radio emission the extensions CoREAS and ZHAireS are enabled.
Both codes have in common that no assumptions on the actual radio emission mechanism are made, but the radiation is calculated by pure electrodynamics applied to each particle in the simulation.
However, the used formalism differs.
In CoREAS the ``endpoint formalism'' \cite{endpoint} is used, whereas in ZHAireS the ``ZHS Algorithm''\cite{ZHS-Algorithm} is implemented.

For the high energetic interaction model, SIBYLL 2.1 is used as it is the best option supported by both codes.
Both take the Earth's curvature into account.
The same atmospheric model is used, namely Linsley's parametrization of the US standard atmosphere.
For the the scaling of the air refractivity $n-1$ to higher altitudes CoREAS uses $n-1$ being proportional 
to the air density, while Aires uses a simple exponential scaling.
To eliminate effects of the refractivity model in the comparison, the CoREAS code has been modified to use an ZHAireS-like simple exponential model for the refractivity.

\section{Calculation of the Radiation Energy}
For an efficient determination of the radiation energy only a small number of antennas is sufficient if placed in a specific way.
The computing time increases almost linearly with the number of simulated antennas. Thus, it is unfeasible to place large numbers of antennas to sample the full two-dimensional emission pattern. 
The model used for an efficient extraction of the radiation energy was introduced in \cite{Glaser:2016qso}.

Instead of placing many antennas to sample the full two-dimensional emission pattern, it is sufficient to simulate only on the  $\vec{v} \times (\vec{v} \times \vec{B})$ axis, as here the polarization of the charge excess and the geomagnetic component decouple. Then, the lateral distribution function (LDF) of the two components can be integrated to obtain the energy fluence $f$ via:

\begin{equation}
  \label{eq:Erd}
   E_{\text{RD}} = 2\pi \int_0^\infty \mathrm{d}r \, r \; f(r).
\end{equation}

In the following, 30 antennas are placed on the $\vec{v} \times (\vec{v} \times \vec{B})$ axis such that the complete radio footprint is covered.
The first antenna is placed at \SI{0.5}{\percent} of the expected size of the radio footprint.
The next twelve antennas are equally spaced from \SI{1}{\percent} to \SI{15}{\percent}.
The last 17 antennas are placed with an equal spacing from \SI{20}{\percent} until the maximal size of the radio footprint.
The spacing is denser close to the shower to detect a possible rapid change of the energy fluence and to be sensitive for the different shapes of the lateral distribution function.
For larger distances this is no longer necessary.
Therefore, this approach ensures an adequate sampling of the radio LDF, such that the uncertainty of the numerical integration can be neglected \cite{Glaser:2016qso}.

To study the impact of the used assumptions and the reduction to antennas on the $\vec{v} \times (\vec{v} \times \vec{B})$
axis 250 air showers with a primary energy between \SIrange{e17}{e19}{\eV} following a uniform distribution of the logarithm of the energy are simulated.
The azimuth angle is distributed uniformly between \SI{0}{\degree} and \SI{360}{\degree} and the zenith angle uniformly between \SI{0}{\degree} and \SI{75}{\degree}.
The geomagnetic field is set to an inclination of \SI{-35.7}{\degree} with a field strength of \SI{0.243}{\gauss} which corresponds to the geomagnetic field at the AERA detector. 

The radio antennas are placed in a star-shape pattern, i.e. in \SI{45}{\degree} lines,
in the shower plane using the same distribution as explained above.
The data are interpolated and integrated numerically over the complete shower plane to compute $E_{\text{RD}}$ from the two dimensional-LDF.
Additionally, the radiation energy is computed using the $\vec{v} \times (\vec{v} \times \vec{B})$ axis only.
An overestimation of \SI{1.68}{\percent} (\SI{1.56}{\percent}) is found for CoREAS (ZHAireS).
The bias can be attributed to the two assumptions made as described in \cite{Glaser:2016qso}.
To correct this bias, $E_{\text{RD}}$ will be reduced accordingly for the following analysis.

\section{CoREAS ZHAireS Comparison}
To make the simulations of both programs comparable, the technical parameters are set to similar values as well as possible.
An individual study is made to exclude any influence of the thinning algorithms in the simulations.
In CORSIKA a thinning level of \num{e-5} with optimal weight limitation is used.
In Aires the thinning level is set to \num{e-5} as well, the statistical weight factor is set to \num{0.06}.

\subsection{Charge Excess Fraction}
Since the used method allows to decompose the radiation energy into a geomagnetic and a charge excess part the charge excess fraction $a$ can be studied directly.
The ratio depends on the geomagnetic angle, i.e. the angle between magnetic field and shower direction, $\alpha$, and is defined as
\begin{equation}
    a = \sin\alpha \sqrt{E_{\text{RD}}^{\text{ce}}/E_{\text{RD}}^{\text{geo}}}.
\end{equation}
Here, $E_{\text{RD}}^{\text{ce}}$ and $E_{\text{RD}}^{\text{geo}}$ are the radiation energies originating from the charge-excess and geomagnetic component respectively.
The square root is taken for consistency with previous work where the electric field amplitude was used instead of the radiation energy.

For the analysis, 1000 proton and 1000 iron induced showers are simulated with energies and directions as mentioned above.
For each shower, the radiation energy originating from the geomagnetic emission and the charge excess is calculated independently.
The charge excess fraction depends on the density at the shower maximum  $\rho(X_{\text{max}})$ where most radiation is emitted.
Using the US standard atmosphere after Linsley, $\rho(X_{\text{max}})$ can be calculated from $X_{\text{max}}$ and the zenith angle $\theta$.

An exponential function of the form
\begin{equation}
 a(\rho(X_{\text{max}})) = q_0 + q_1 \cdot \exp{\left( q_2(\rho(X_{\text{max}}) - \rho(\langle X_{\text{max}} \rangle)) \right)}
\end{equation}
is fitted to the data. $\rho(\langle X_{\text{max}} \rangle)  = \SI{0.65}{\kilo\gram\per\meter\squared}$ is the air density at the shower maximum for an average zenith angle of \SI{45}{\degree} and an average $\langle X_{\text{max}} \rangle = \SI{669}{\gram\per\centi\meter\cubed}$ as predicted by QGSJETII-04 for a shower energy of \SI{1}{\EeV} and a \SI{50}{\percent} proton/\SI{50}{\percent} iron composition \cite{1475-7516-2013-07-050}. 

\begin{figure}
\begin{subfigure}[c]{\linewidth}
\centering
\includegraphics[width=.9\linewidth]{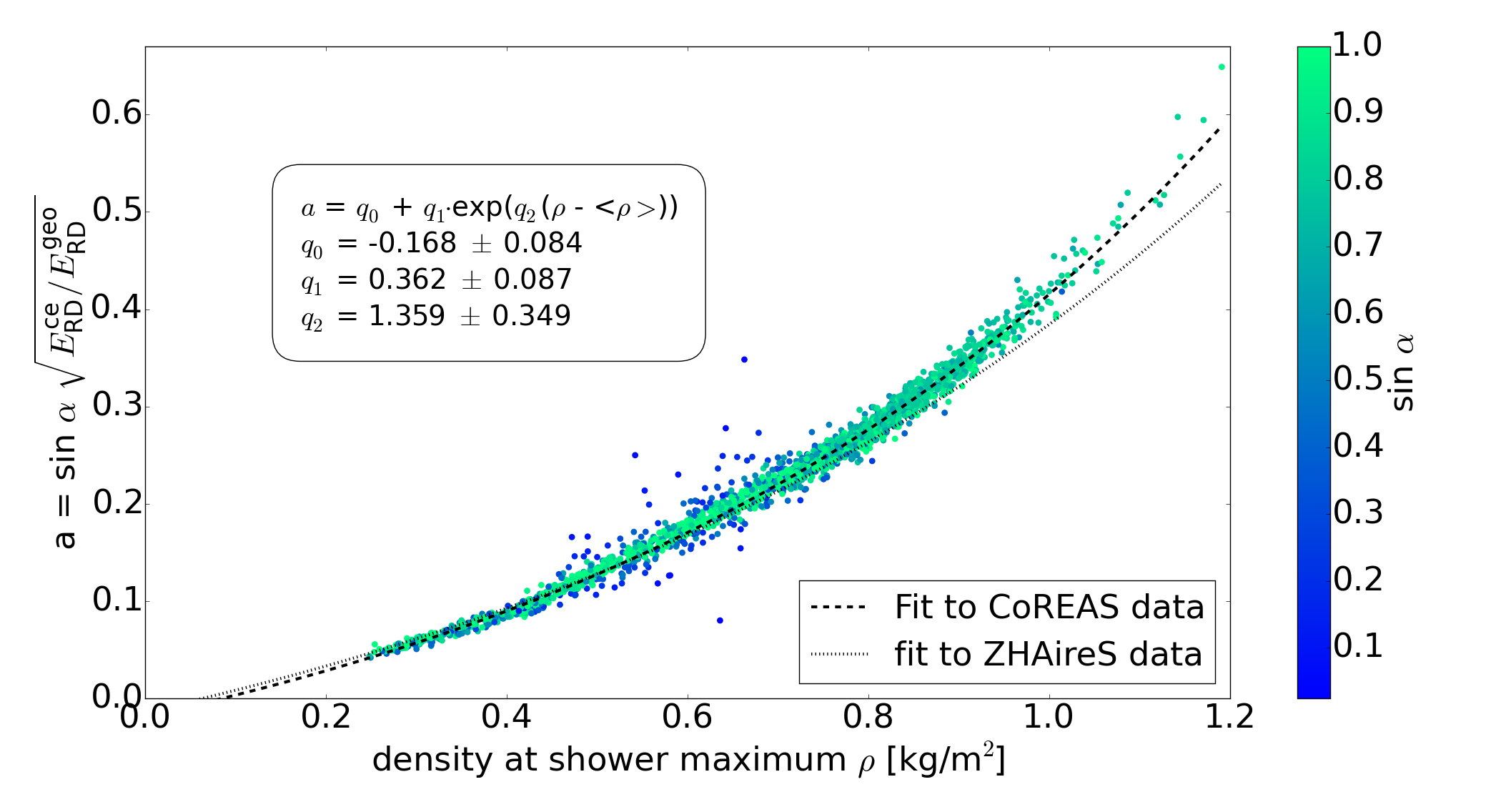}
\end{subfigure}
\begin{subfigure}[c]{\linewidth}
\centering
\includegraphics[width=.9\linewidth]{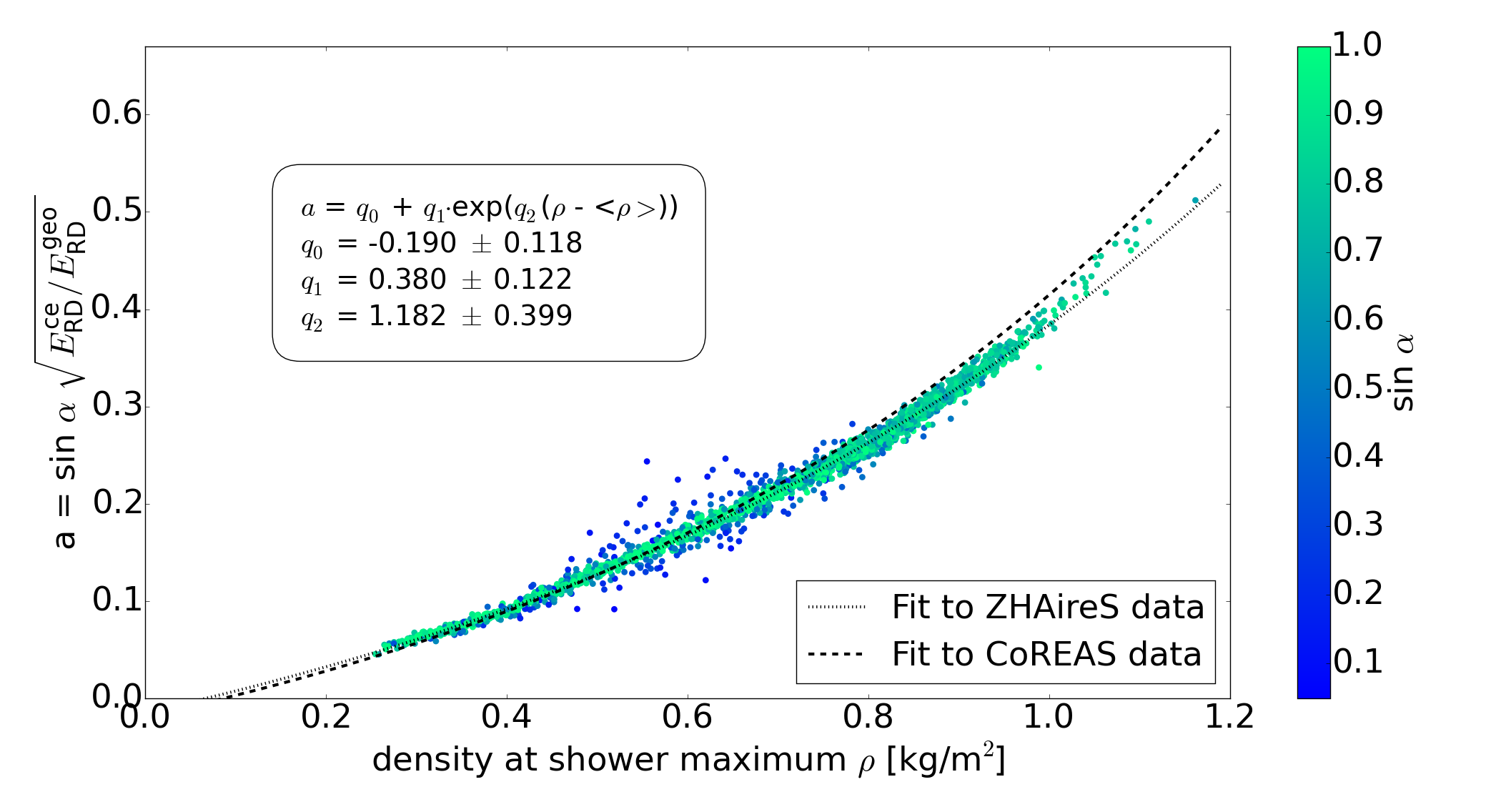}
\end{subfigure}
\caption{Charge excess fraction $a$ of air showers depending on the atmospheric density at the shower maximum for CoREAS (top) and ZHAireS (bottom). The fit of the other code is added in both plots for a direct comparison.}
\label{fig:a}
\end{figure} 

In Fig. \ref{fig:a}, the charge excess fraction including the fit results for both programs is shown\footnote{Due to missing uncertainties on the data the absolute scale of the uncertainties on the fit parameters is arbitrary - 
the uncertainties only allow to judge the relative constraints on the fit parameters.}. 
The colour code indicates that most outliers are air showers with a small $\sin\alpha$ value. 
Comparing the two fitted functions shows a good agreement between CoREAS and ZHAireS. 
A deviation is found for showers with a high density at the shower maximum which corresponds to showers with small zenith angles.

\subsection{Radiation Energy}
As the radiation is almost solely emitted by the electromagnetic part of the air shower, it correlates best with the energy of the electromagnetic cascade. Due to the much lower charge to mass ratio muons hardly emit radiation and can be neglected.
A method to express the invisible energy as a function of the electromagnetic energy is presented in \cite{ThePierreAuger:2013eja}. Combining both allows to compute the primary cosmic-ray energy from the radiated energy of the electromagnetic cascade.

Before the correlation with the electromagnetic energy, the radiation energy has to be corrected for various effects.
In a first step the geomagnetic part of the emission is corrected for the shower direction as it scales with $\sin^2\alpha$. 
A second correction arises due to differences in the $X_{\text{max}}$ values. A detailed explanation of these corrections can be found in \cite{Glaser:2016qso}. 
The final corrected radiation energy with two free parameters $p_0$ and $p_1$ is then given by

\begin{equation}
 S_{\text{RD}} = \frac{E_{\text{RD}}}{a(\rho(X_{\text{max}}))^2 + (1-a(\rho(X_{\text{max}}))^2)\sin^2\alpha} \cdot \frac{1}{(1-p_0+p_0\exp{[p_1(\rho(X_{\text{max}})-\rho(\langle X_{\text{max}} \rangle))]})^2}
\end{equation}

\begin{figure}
\centering
\begin{subfigure}[c]{\linewidth}
\centering
\includegraphics[width=.9\linewidth]{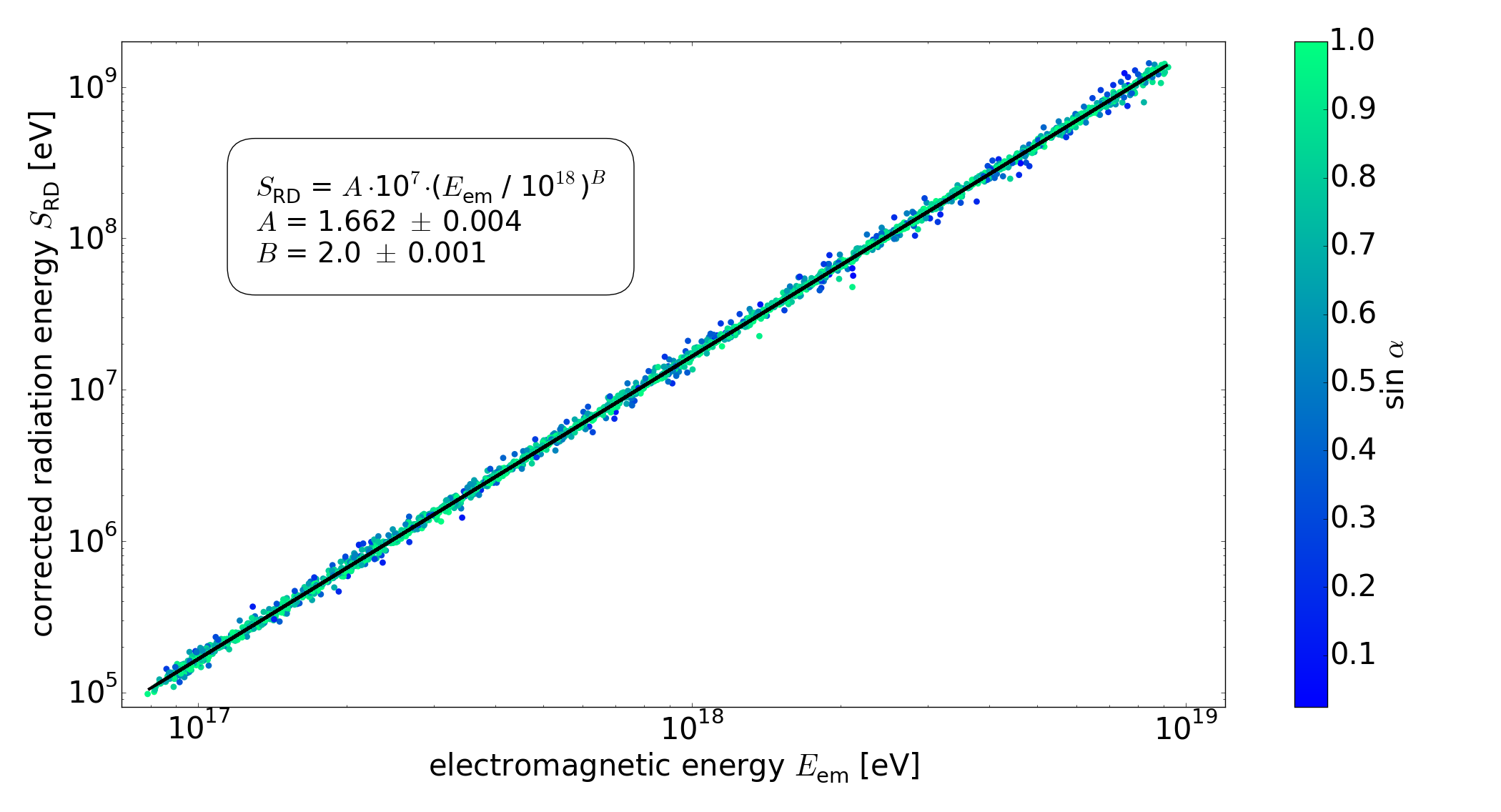}
\end{subfigure}
\begin{subfigure}[c]{\linewidth}
\centering
\includegraphics[width=.9\linewidth]{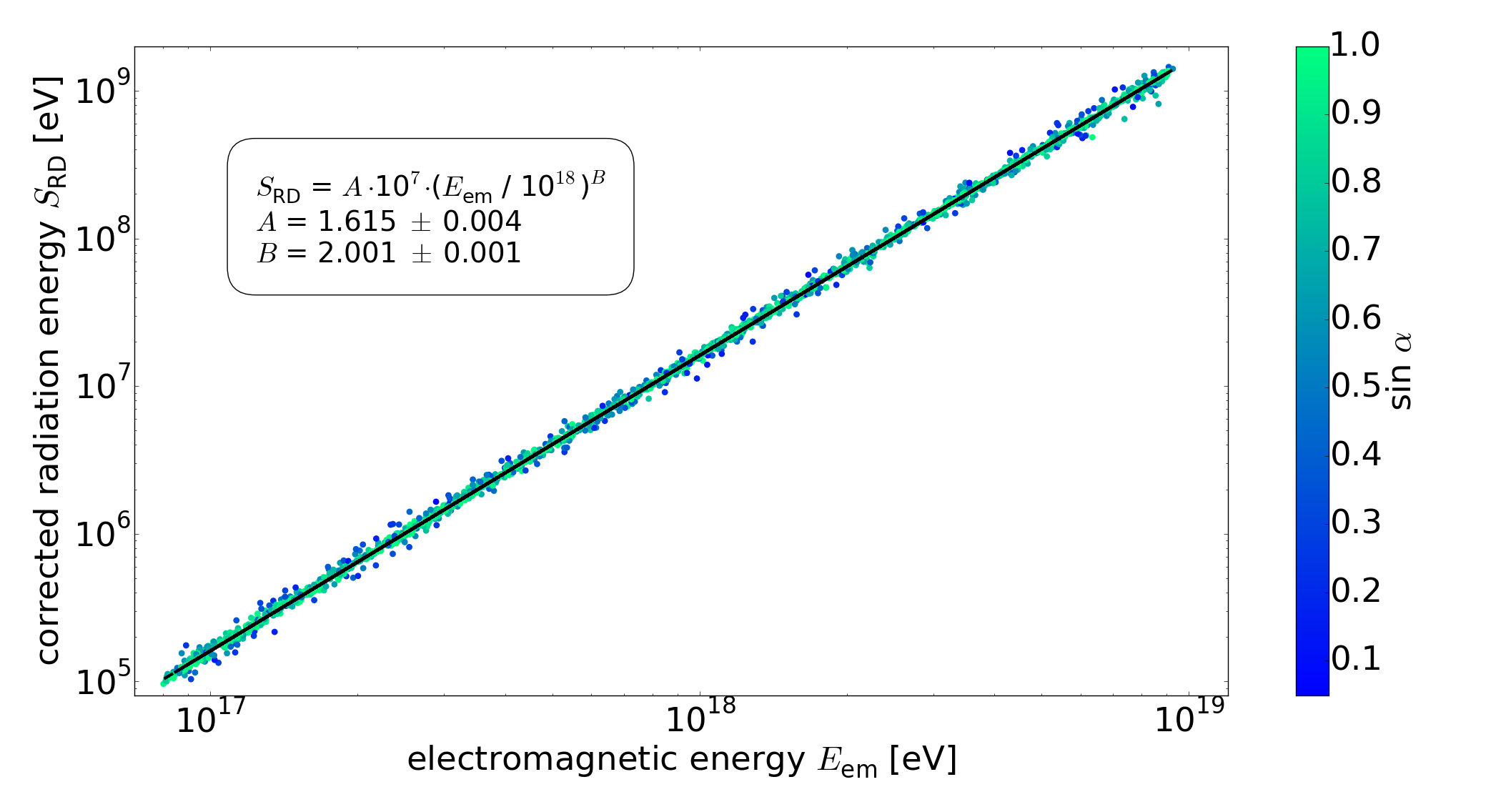}
\end{subfigure}
\caption{Correlation between the corrected radiation energy and the electromagnetic component of air showers for CoREAS (top) and ZHAireS (bottom).}
\label{fig:S}
\end{figure}

The parameters $p_0$ and $p_1$ are determined in a combined fit with the power law
\begin{equation}
\label{eq:PowerLaw}
 S_{\text{RD}} = A \cdot \SI{e7}{\eV} (E_{\text{em}}/\SI{e18}{\eV})^B.
\end{equation}
The results are given in table \ref{tab:fitparam}, the correlations between the corrected radiation energy and the electromagnetic energy are shown in Fig.~\ref{fig:S}. 
The slope $B$ is exactly equal to two as it is expected for a coherent emission. 
Looking at the deviation of data and fit a scatter of roughly \SI{7}{\percent} is found.
The difference between the CoREAS and ZHAireS prediction is given by the ratio of $A_\text{CoREAS}/A_\text{ZHAireS} = \SI{2.9}{\percent}$ 
which corresponds to a difference in the electromagnetic energy of \SI{1.4}{\percent}.

\begin{table}
\centering
\begin{tabular}{r
   S[table-format=-1.3,table-figures-uncertainty=1]
   S[table-format=-1.3,table-figures-uncertainty=1]}
 \toprule
   & {CoREAS} & {ZHAireS}\\
 \midrule
 $A$ & 1.662(4) & 1.615(4) \\
 $B$ & 2.000(1) & 2.001(1) \\
 $p_0$ & 0.293(7) & 0.291(7) \\
 $p_1$ & -2.781(55) & -2.739(60) \\
 \bottomrule
\end{tabular}
\caption{Best fit parameters for CoREAS and ZHAireS using 1000 proton and 1000 iron induced showers.}
\label{tab:fitparam}
\end{table}

The fits are repeated on a combined set of the data from both codes to obtain a single set of parameters $q_i$ and $p_i$.
Using them the differences between CoREAS and ZHAireS are not reduced by applying individual corrections for each code.
Fitting the power law with the combined corrections for each code individually a difference of \SI{3.3}{\percent} is observed. This corresponds to a difference in the electromagnetic energy of \SI{1.6}{\percent}.
Details of the fit are given in table \ref{tab:fitparam_comb}, the individual and the combined parameters are summarized in table \ref{tab:combined}.

\begin{table}
\centering
\begin{tabular}{r
   S[table-format=-1.3,table-figures-uncertainty=1]
   S[table-format=-1.3,table-figures-uncertainty=1]
   S[table-format=-1.3,table-figures-uncertainty=1]}
 \toprule
   & {CoREAS} & {ZHAireS} & {combined}\\
 \midrule
    $q_0$ & -0.168(84) & -0.190(118) & -0.184(72)\\
    $q_1$ & 0.362(87) & 0.380(122) & 0.376(74)\\
    $q_2$ & 1.359(349) & 1.182(399) & 1.250(263)\\
    $p_0$ & 0.293(7)& 0.291(7) & 0.290(5)\\
    $p_1$ & -2.781(55) & -2.739(60) & -2.774(42)\\
 \bottomrule
\end{tabular}
\caption{Individual and combined parameters for the corrections of the radiation energy. The $q_i$ values belong to the charge excess fraction, $p_i$ are used in the second correction term.}
\label{tab:combined}
\end{table}

\begin{table}
\centering
\begin{tabular}{r
   S[table-format=-1.4,table-figures-uncertainty=1]
   S[table-format=-1.4,table-figures-uncertainty=1]}
 \toprule
   & {CoREAS} & {ZHAireS}\\
 \midrule
 $A$ & 1.6635(11) & 1.6105(11) \\
 $B$ & 2.0002(5) & 2.0007(5) \\
 \bottomrule
\end{tabular}
\caption{Best fit parameters for CoREAS and ZHAireS with the combined corrections using 1000 proton and 1000 iron induced air showers.}
\label{tab:fitparam_comb}
\end{table}

These results have been obtained with the standard settings of the two codes regarding the path length between interactions. Recently we have observed hints that the predicted radiation energy increases slightly when the particle cascade is simulated more finely. We are currently investigating this effect in a direct comparison between the two codes.

\subsection{Additional Checks}
Additional checks are done to validate the found agreement. Varying the refractive index at sea level and the geomagnetic field strength have shown no significant difference between CoREAS and ZHAireS. 
Differences between proton and iron induced air showers are analysed as well. Both programs predict around \SI{3}{\percent} more radiation for a proton induced air shower than for an iron one with the same electromagnetic energy.

A last check is performed directly on the lateral distribution function. In the shower plane the LDF can be described as \cite{Aab:2015vta}
\begin{equation}
\label{eq:LDF2}
 f(\vec{r}) = A \left( \exp\left(-\frac{\vec{r}+C_1\vec{e}_{\vec{v} \times \vec{B}}-\vec{r}_{\text{core}}}{\sigma^2}\right) - C_0 \exp\left(-\frac{\vec{r}+C_2\vec{e}_{\vec{v} \times \vec{B}}-\vec{r}_{\text{core}}}{(C_3\exp(C_4\cdot\sigma))^2}\right) \right) ,
\end{equation}
where $\vec{r}$ denotes the station position, $\vec{r}_{\text{core}}$ the position of the core in the shower plane and constants $C_i$ that are obtained by simulations and depend on the zenith angle. 
The comparison of CoREAS and ZHAireS have shown consistent results for the mean values per zenith bin for each constant.

\section{Conclusion}
The radio emission of extensive air showers has been studied with CoREAS and ZHAireS simulations. 
The technical parameters in the simulations were set to similar values as well as possible. 
1000 proton and 1000 iron induced air showers have been simulated to study the differences between CoREAS and ZHAireS. 

The used method allows to decompose the radiation into the part originating from the geomagnetic emission and the charge excess. 
Their ratio, the charge excess fraction, depends on the density at the shower maximum as most of the radiation is emitted close to it. 
The charge excess fraction can be described with an exponential function. 
Only for showers with a high density at the shower maximum the simulations have shown some deviation.

After correcting the radiation energy for the geomagnetic angle and density at the shower maximum 
it correlates with the electromagnetic energy of the air shower. The correlation can be expressed as a quadratic power law.
We find a good agreement between the CoREAS and ZHAireS code. The absolute prediction of the radiation energy for given electromagnetic energy agrees within \SI{3.3}{\percent}.
We note that a thorough analysis of the step sizes used in the two codes, which might influence this result, still needs to be performed.
If this difference is interpreted as a systematic uncertainty on the absolute prediction, this amounts to \SI{1.6}{\percent} uncertainty of the electromagnetic energy, well below current experimental limits.

%

\end{document}